\documentclass[lettersize,onecolumn]{IEEEtran}
\usepackage{amsmath,amssymb,amsthm}
\usepackage{array}
\usepackage{url}
\usepackage{cite}
\usepackage{textcomp}
\newtheorem{theorem}{Theorem}
\newtheorem{lemma}{Lemma}

\newcommand{\Z}{\mathbb{Z}}
\newcommand{\tS}{\widetilde{S}}

\begin{document}

\title{Comments on ``Multidimensional Costas Arrays and Their Periodicity''}

\author{Shiv~Munagala%
\thanks{Manuscript submitted October 2026.}%
\thanks{S.~Munagala is an independent researcher, London, U.K. (e-mail: \texttt{contact@shivm.xyz}).}}

%\markboth{IEEE TRANSACTIONS ON INFORMATION THEORY}%
%{Munagala: Comments on ``Multidimensional Costas Arrays and Their Periodicity''}

\maketitle

\begin{abstract}
Rubio and Torres conjectured that a multidimensional Costas array, defined by a bijection between two boxes of integer points whose domain has at least as many dimensions as its image, can stay Costas in every window of its periodic extension only if its order is a power of two fixed by the dimension of the domain. We give a four-dimensional periodic Costas array of order $6$, with a two-dimensional domain and a two-dimensional image, so the conjecture is false. The verification reduces to $36$ windows. It has been confirmed by computer and in the Lean proof assistant. The array also meets with equality the pigeonhole bound that Rubio and Torres use for their non-existence results, so that bound is tight.
\end{abstract}


\section{Definitions}
Write $[a]=\{1,\dots,a\}$. Following \cite[Defs.~2--4]{ref1}, let $X=[n_1]\times\cdots\times[n_k]$ and $Y=[n_{k+1}]\times\cdots\times[n_m]$ with all $n_i\ge 2$, and let $\varphi:X\to Y$ be a bijection. Its set of dots is $S=\{(x,\varphi(x)):x\in X\}\subset\Z^m$, and $n=|X|$ is the order. Let $\Lambda=[n_1]\times\cdots\times[n_m]$ and let $\tS=S+(n_1\Z\times\cdots\times n_m\Z)$ be the periodic extension of $S$.

The array is \emph{Costas} if the difference vectors $\omega-\alpha$, over ordered pairs of distinct dots $\alpha,\omega\in S$, are pairwise distinct. It is \emph{periodic Costas} if it is Costas and, for every $t\in\Z^m$, the dots in the window $\tS\cap(t+\Lambda)$ have pairwise distinct difference vectors. These are ordinary integer differences, not reduced modulo the $n_i$.

\medskip
\noindent\textbf{Conjecture} \cite[Conj.~1]{ref1}. \emph{If $k\ge m-k$ and the array is periodic Costas, then $n=2^k$, that is, $n_1=\cdots=n_k=2$.}

\medskip
This is true for $m=2$, which is a theorem of Taylor \cite{taylor} (see \cite[Thm.~1 and Remark~2]{ref1}), and for $m=3$ \cite[Thm.~5]{ref1}. It is also listed as AIM problem 627 \cite{ref2}.

\section{A Counterexample}
Take $m=4$, $k=2$ and $X=Y=[2]\times[3]$, so that $n=6\neq 4=2^2$. Define $\varphi$ by
\begin{center}
\footnotesize
\setlength{\tabcolsep}{3pt}
\begin{tabular}{c|cccccc}
$x$ & $(1,1)$ & $(1,2)$ & $(1,3)$ & $(2,1)$ & $(2,2)$ & $(2,3)$\\\hline
$\varphi(x)$ & $(1,1)$ & $(1,2)$ & $(2,1)$ & $(1,3)$ & $(2,3)$ & $(2,2)$
\end{tabular}
\end{center}
so that
$$S=\{(1,1,1,1),(1,2,1,2),(1,3,2,1),(2,1,1,3),(2,2,2,3),(2,3,2,2)\}.$$

\begin{theorem}\label{thm:main}
$S$ is a periodic Costas array of order $6$. Hence \cite[Conj.~1]{ref1} is false.
\end{theorem}

\begin{lemma}\label{lem:res}
Let $p=(2,3,2,3)$. For every $t\in\Z^4$, the window at $t$ is a translate of the window at $\rho(t)=(t_i \bmod p_i)_i\in\prod_i\{0,\dots,p_i-1\}$.
\end{lemma}

\begin{IEEEproof}
Write $t=\rho(t)+u$ with $u\in p_1\Z\times\cdots\times p_4\Z$. Since $\tS+u=\tS$, we have $\tS\cap(t+\Lambda)=\bigl(\tS\cap(\rho(t)+\Lambda)\bigr)+u$.
\end{IEEEproof}

\begin{IEEEproof}[Proof of Theorem~\ref{thm:main}]
The table shows that $\varphi$ is a bijection. Distinctness of difference vectors is invariant under translation, so by Lemma~\ref{lem:res} it suffices to check the $2\cdot3\cdot2\cdot3=36$ windows with $t\in\prod_i\{0,\dots,p_i-1\}$. Since each side of $\Lambda$ equals the corresponding period $p_i$, every window contains exactly one translate of each dot of $S$, hence exactly six dots, and a direct computation shows that its $30$ difference vectors are distinct. The window $t=0$ is $S$ itself, so $S$ is also Costas.
\end{IEEEproof}

\emph{Verification.} The computation above is a short script, available as supplementary material, which checks the bijection, the Costas property, all $36$ residue windows, and, independently of Lemma~\ref{lem:res}, every offset $t\in\{-6,\dots,6\}^4$. The statement has also been formalised and checked in Lean~4 using kernel evaluation (\texttt{decide +kernel}), depending only on the axioms \texttt{propext}, \texttt{Classical.choice} and \texttt{Quot.sound} \cite{ref3}. The formal definitions follow \cite[Defs.~2--4]{ref1}: one-based boxes, integer differences, and all offsets $t\in\Z^m$. A preliminary note containing the same example was posted as a preprint \cite{ref4}.

\section{Tightness of Theorem 3 of \cite{ref1}}
For a permutation array $A$ of order $n$, \cite[Thm.~3]{ref1} states that if $|\mathcal{H}_A|-|H_\Lambda|<n-1$, then some window of the periodic extension has a repeated difference vector. Here, following \cite[Defs.~1 and~6]{ref1}, the toroidal vector of an ordered pair of dots is their difference reduced modulo $(n_1,\dots,n_m)$; $T_\Lambda$ is the set of the $(n-1)^2$ toroidal vectors that can occur between two distinct dots of a permutation array, namely those that are nonzero in both the domain and the image coordinates; $H_\Lambda\subseteq T_\Lambda$ is the subset having some component $h_i=n_i/2$; and $\mathcal{H}_A$ is the multiset of toroidal vectors of $A$ lying in $H_\Lambda$.

For the array above, of the $n(n-1)=30$ toroidal vectors, $|\mathcal{H}_A|=26$ have a component equal to $n_i/2$ (the even sides are $n_1=n_3=2$), while $|H_\Lambda|=21$ of the $|T_\Lambda|=25$ admissible toroidal vectors do. Hence
\[
|\mathcal{H}_A|-|H_\Lambda| = 5 = n-1,
\]
so the hypothesis of \cite[Thm.~3]{ref1} fails with equality. Repeated toroidal vectors do occur (for example $\langle 1,1,1,2\rangle$ occurs three times), but all of them lie in $\mathcal{H}_A$, where \cite[Prop.~1]{ref1} does not apply. Thus the pigeonhole bound in \cite[Thm.~3]{ref1} is tight and cannot be sharpened by the same argument.

\section{Remarks}
The counterexample does not affect the non-existence results of \cite{ref1} for odd order \cite[Cor.~2]{ref1}, for arrays whose domain or image is one-dimensional under the hypothesis of \cite[Thm.~4]{ref1}, or for $m=3$ \cite[Thm.~5]{ref1}. Since $m=4$ is the smallest dimension not covered by those results, the order-$2^k$ conclusion fails as soon as the conjecture goes beyond the cases proved in \cite{ref1}. Whether the example belongs to a wider family of periodic Costas arrays of order other than $2^k$ is left open.

\section*{Acknowledgments}
The author thanks J.~Torres for confirming that the definition of periodic Costas used here matches that of \cite{ref1} and for observing that the example attains equality in \cite[Thm.~3]{ref1}.

\emph{Disclosure of AI use.} The counterexample was found by computer search carried out with the assistance of Claude (Anthropic). Claude also wrote the verification script and the Lean~4 proof described in Section~II, and drafted the text of this Comment, which the author reviewed, edited and takes full responsibility for. The mathematical claims rest on the elementary argument in Section~II, which the author has checked by hand, and on the Lean proof, which was checked by the proof assistant.


\begin{thebibliography}{1}
\bibitem{ref1}
I.~Rubio and J.~Torres, ``Multidimensional Costas arrays and their periodicity,'' \emph{IEEE Trans. Inf. Theory}, vol.~69, no.~8, pp.~5032--5040, Aug. 2023, doi: 10.1109/TIT.2023.3264951.

\bibitem{taylor}
H.~Taylor, ``Non-attacking rooks with distinct differences,'' Commun. Sci. Inst., Univ. Southern California, Los Angeles, CA, USA, Tech. Rep. CSI-84-03-02, 1984.

\bibitem{ref2}
``Problem 627: Periodic Costas arrays,'' AIM Open Applied Problems. Accessed: Oct. 9, 2026. [Online]. Available: \url{https://github.com/MColbrook/AIM/blob/c9929805cece044705abaaebe1cebddac1ba11a6/problems/627-periodic-costas.md}

\bibitem{ref3}
S.~Munagala, ``Periodic multidimensional Costas arrays (Rubio--Torres Conjecture~1),'' Prove2Me. Accessed: Oct. 9, 2026. [Online]. Available: \url{https://prove2.me/missions/Periodic_Multidimensional_Costas_Arrays_(Rubio%E2%80%93Torres_Conjecture_1)}

\bibitem{ref4}
S.~Munagala, ``A periodic Costas array of order 6: A counterexample to the Rubio--Torres conjecture,'' Zenodo, Oct. 2026, doi: 10.5281/zenodo.23209356.
\end{thebibliography}
\end{document}